\documentclass[prl,aps,amssymb,twocolumn,superscriptaddress,notitlepage]{revtex4-1}
\usepackage{amsmath}
\usepackage{amssymb}
\usepackage{amsthm}
\usepackage{amsfonts}
\usepackage{listings}
\usepackage{enumerate}
\usepackage{latexsym}
\usepackage{psfrag}
\usepackage{bm}
\usepackage{graphicx}
\usepackage[caption=false]{subfig}
\usepackage{blkarray}
\usepackage{array}
\usepackage{color}
\usepackage[normalem]{ulem}
\usepackage{hyperref}

\def\half{\frac{1}{2}}

\begin{document}
\title{Revisiting (higher order and crystalline) topology in old models of lead telluride}
\author{I\~{n}igo Robredo}
\affiliation{Donostia International Physics Center, 20018 Donostia-San Sebastian, Spain}
\affiliation{Department of Condensed Matter Physics, University of the Basque Country UPV/EHU, Apartado 644, 48080 Bilbao, Spain}
\author{Maia G. Vergniory}
\affiliation{Donostia International Physics Center, 20018 Donostia-San Sebastian, Spain}
\affiliation{IKERBASQUE, Basque Foundation for Science, Maria Diaz de Haro 3, 48013 Bilbao, Spain}
\author{Barry Bradlyn}%
\affiliation{Department of Physics and Institute for Condensed Matter Theory, University of Illinois at Urbana-Champaign, Urbana, IL, 61801-3080, USA}%
\date{\today}
\begin{abstract}
In this work, we revisit the model of PbTe presented in Ref.~\onlinecite{fradkin1}. We show that the low energy theory of this model corresponds to a (higher-order) topological crystalline insulator in space group $Fm\bar{3}m1'$, diagnosable by symmetry indicators. We show that the gapless fermions found on antiphase domain walls are the topological boundary modes of the system, due to a nonvanishing mirror Chern number. Furthermore, we show that any symmetric completion of the model must be in this same topological phase. Finally, we comment on the relationship of this model to realistic PbTe, which has recently been predicted to have a phase which realizes same bulk symmetry indicators.
\end{abstract}

\maketitle
\paragraph{Introduction.}One of the most striking features of topological insulators is the presence of protected gapless modes at surfaces, interfaces, and defects\cite{ssh1979,teo2010topological}. The best-known examples of this are the helical modes at the boundary of a two-dimensional ``strong'' topological insulator\cite{Kane04,bernevig2006quantum,konig2007quantum}, and the single Dirac fermion at the two-dimensional boundary of a three-dimensional ``strong'' topological insulator\cite{fukanemele,xia2009observation}. Both of these surface modes are protected by time-reversal symmetry alone. It has recently been appreciated that, with additional crystalline symmetries, more exotic topologically protected boundary features may emerge, such as the multiple Dirac fermions at symmetric boundaries of a mirror Chern insulator\cite{Teo08,Hsieh2012,Tanaka2012}. Even more surprising, ``higher-order TIs'' with (roto)-inversion symmetries may feature topologically protected ``hinge'' modes which propagate on boundaries of two (or more) fewer dimensions than the bulk\cite{Fu2011,Wieder17,fangrotational,benalcazar2017electric,benalcazar2017quantized,hotis,Po2017,Kruthoff2016,NaturePaper,khalaf,song2017}.

In light of these recent developments, we may be tempted to take a fresh look at the old observation that a simplified tight-binding model for PbTe was found to host four Dirac fermions on a two-dimensional antiphase boundary\cite{fradkin1,fradkin2,fradkin3}. It was realized early on that this effective model did not quite respect the symmetries of the crystal\cite{wilczekaxion,haldanemodel}, and the even number of Dirac cones disqualified this model from being a strong TI regardless\cite{Tchernyshyov2000,Fu2007}. Furthermore, realistic PbTe is known to have zero mirror Chern number\cite{Hsieh2012}, and nevertheless the effective model does not possess the requisite mirror symmetry. It is natural to ask, then: are the domain wall fermions of Refs.~\onlinecite{fradkin1,fradkin2,fradkin3} a signifier of any topological crystalline insulating (TCI) phase?

In this work, we answer this question in the affirmative. First, we review the model of Ref.~\onlinecite{fradkin1} (hereafter referred to as the FDB model), and show how to modify it in order to respect the symmetries of the cubic space group $Fm\bar{3}m1'$ (225)\footnote{the final $1'$ denotes the presence of time-reversal symmetry\cite{ITA}}. In particular, we will show that the low-energy effective models obtained from either the FDB model or our symmetric model are unitarily equivalent. Next, we will show that our improved tight-binding model captures the transition between a trivial insulator and a mirror Chern and higher-order topological insulator (HOTI), with symmetry-indicated topological index $\nu=4\in\mathbb{Z}_8$. Because the dynamics of this transition are captured by the low energy effective model, we will show that the four Dirac cones in the FDB model at an antiphase domain wall are the boundary modes of this mirror and higher-order topological insulator. We will show that any symmetric completion of the FDB model has $\nu=4$, with mirror Chern number $\nu_{m_{1\bar{1}0}}=2$ and higher-order ``$S_4$'' invariant $\delta_{S_4}=1$\cite{khalaf,song2017,Po2017,bigmaterials} (Here $m_{1\bar{1}0}$ denotes the mirror about $\mathbf{\hat{x}}-\mathbf{\hat{y}}$, and $S_4$ signifies a fourfold roto-inversion). Finally, we will connect these results to the recent prediction\cite{bigmaterials,bigmaterials-china,bigmaterials-ashvin} that PbTe may be, in some cases, a TCI/HOTI. In doing so, we will see the importance of careful structural determination for finding small-gap topological materials.

\paragraph{Effective tight-binding model} Let us start by reviewing the FDB model, exploring its shortcomings, and constructing an improved model with the same phenomenology. Let us start as did the authors of Ref.~\onlinecite{fradkin1} by noting that PbTe has a rocksalt structure, with the symmetries of space group $Fm\bar{3}m1'$ (225), the centrosymmetric, symmorphic space group with a face-centered cubic Bravais lattice and octahedral point group. We take as a basis for the Bravais lattice (we set the lattice constant $a=1$ for simplicity)
\begin{equation}
\mathbf{t}_1=\frac{1}{2}(\hat{\mathbf{y}}+\hat{\mathbf{z}}),\;\;\mathbf{t}_2=\frac{1}{2}(\hat{\mathbf{x}}+\hat{\mathbf{z}}),\;\;\mathbf{t}_3=\frac{1}{2}(\hat{\mathbf{x}}+\hat{\mathbf{y}}).
\end{equation}
The rocksalt structure has Te atoms located at the $4a$ Wyckoff position, with reduced coordinates $\mathbf{q}_a=(0,0,0)$, and Pb atoms at the $4b$ Wyckoff position with reduced coordinates $\mathbf{q}_b=(1/2,1/2,1/2)$. The point group of PbTe is generated by a threefold rotation $C_{3,111}$ about the body diagonal of the unit cell, a fourfold rotation $C_{4x}$ about the $\mathbf{\hat{x}}=\mathbf{t}_2+\mathbf{t_3}-\mathbf{t}_1$ axis, spatial inversion $I$, and time-reversal symmetry $T$. The original model of FDB consisted of spin-1/2 s orbitals on the $4a$ and $4b$ Wyckoff positions as a proxy for the Te and Pb atoms (whether we start from s or p orbitals, the sign of the inversion matrix will change, but all topological properties remain invariant). They added a staggered on-site potential taking opposite values $\pm m$ on the Pb and Te sites, and a nearest-neighbor spin-dependent hopping. Letting $\vec{\tau}$ be a set of Pauli matrices acting in the orbital (Te,Pb) basis, and letting $\vec{\sigma}$ be the set of Pauli matrices acting on spin, we can write the Bloch Hamiltonian for the FDB model as
\begin{equation}
H_{\mathrm{FDB}}=m\tau_z+t\tau_y\sum_{\mu=x,y,z}\sigma_\mu\cos\left(\frac{k_\mu}{2}\right),\label{eq:fdb}
\end{equation}
where we have taken the liberty of restoring a neglected factor of $i$ in the hopping term to restore time-reversal symmetry\cite{haldanemodel}\footnote{While this makes our Eq.~(\ref{eq:fdb}) formally different from the original FDB model, we view the lack of time-reversal symmetry in Ref.~\cite{fradkin1} to be a simple oversight.} Eq.~(\ref{eq:fdb}) is manifestly time-reversal, inversion, and $C_{3,111}$ symmetric; each of these symmetries acts trivially in the orbital $\vec{\tau}$ space, and as a rotation (the identity rotation for inversion) on the spin degrees of freedom. Precisely, we have for these symmetries $\{g\}$ that
$\Delta(g)^{-1}H_{\mathrm{FDB}(\mathbf{k})}\Delta(g)=H_{\mathrm{FDB}}(g\mathbf{k})$,
with the matrix representation $\Delta$ specified by
\begin{align}
\Delta(C_{3,111})&=\exp[\frac{-i\pi}{3\sqrt{3}}(\sigma_x+\sigma_y+\sigma_z)],\label{eq:delta1}\\
\Delta(I)&=\mathbb{I}_{4\times4},\\
\Delta(T)&=i\sigma_y\mathcal{K}.\label{eq:delta2}
\end{align}
The spectrum of this Hamiltonian consists of two sets of doubly-degenerate bands (due to $IT$ symmetry) separated by a spectral gap which is smallest at the $L$ point $(1/2,1/2,1/2)$ and given by $\delta E_L=2m$. Depending on the sign of $m$, there is a band inversion at the $L$ point: for $m>0$ the valence bands carry the representation $\bar{L}_9$ of the little group of $L$ (with inversion eigenvalues $(-1,-1)$), while for $m<0$ the valence bands carry the representation $\bar{L}_8$ of the little group of $L$ (with inversion eigenvalues $(+1,+1)$). Because there are four $L$ points in the FCC Brillouin zone, this is not a $\mathbb{Z}_2$ nontrivial TI\cite{Fu2007}, and so we must look for nontrivial TCI invariants.

Here, however, we run into a problem: The Hamiltonian $H_{\mathrm{FDB}}$ is $C_{4x}$-symmetric, but the matrix representative of $C_{4x}$ can be seen to be
\begin{equation}
\Delta(C_{4x})=\tau_z\sigma_y\exp(-i\pi/4\sigma_x).
\end{equation}
This has the unfortunate property that $\Delta(C_{4x})^4=+1$, rather than $-1$ as needed for a double-valued representation. Repairing this by multiplying by a factor of $\sqrt{i}$ is futile, as then $\Delta(C_{4x})$ and $\Delta(T)$ no longer commute. Thus, the FDB Hamiltonian does not have the symmetries of the space group $Fm\bar{3}m1'$. 

\begin{figure}[t]
\subfloat[]{
	\includegraphics[width=0.2\textwidth]{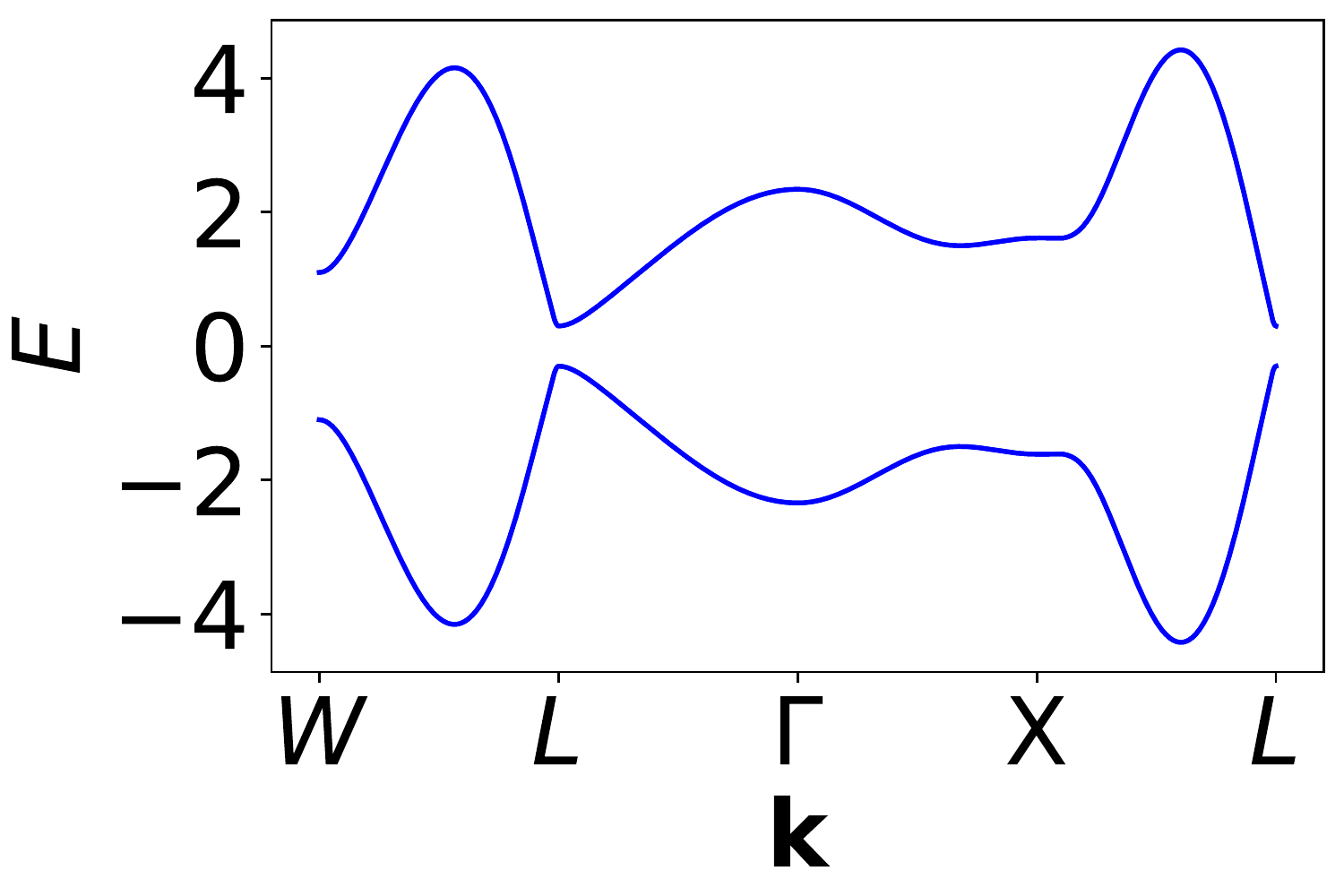}\label{fig:modelbulkspec1}
}
\subfloat[]{
	\includegraphics[width=0.2\textwidth]{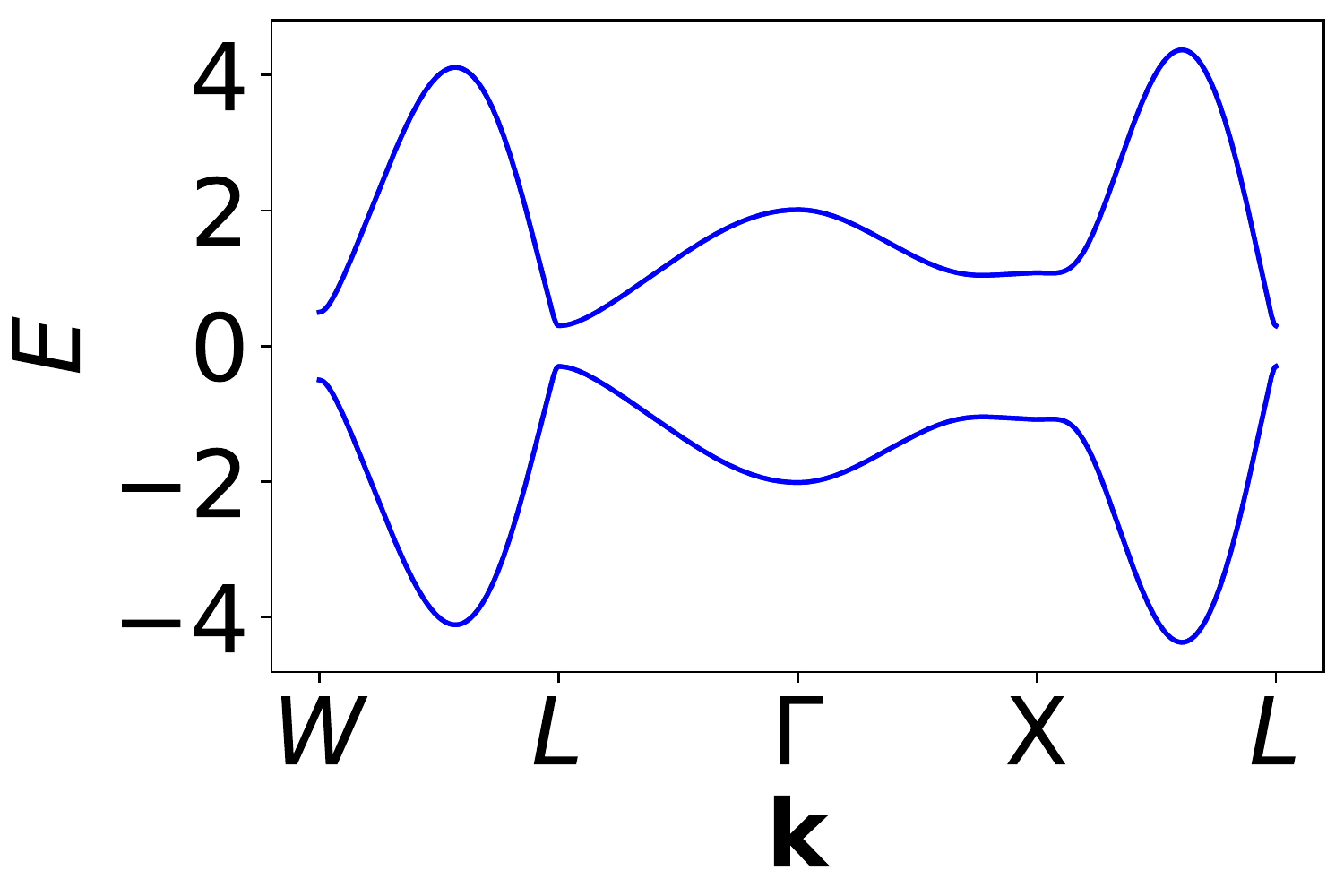}\label{fig:modelbulkspec2}
}\qquad
\subfloat[]{
    \includegraphics[width=0.2\textwidth]{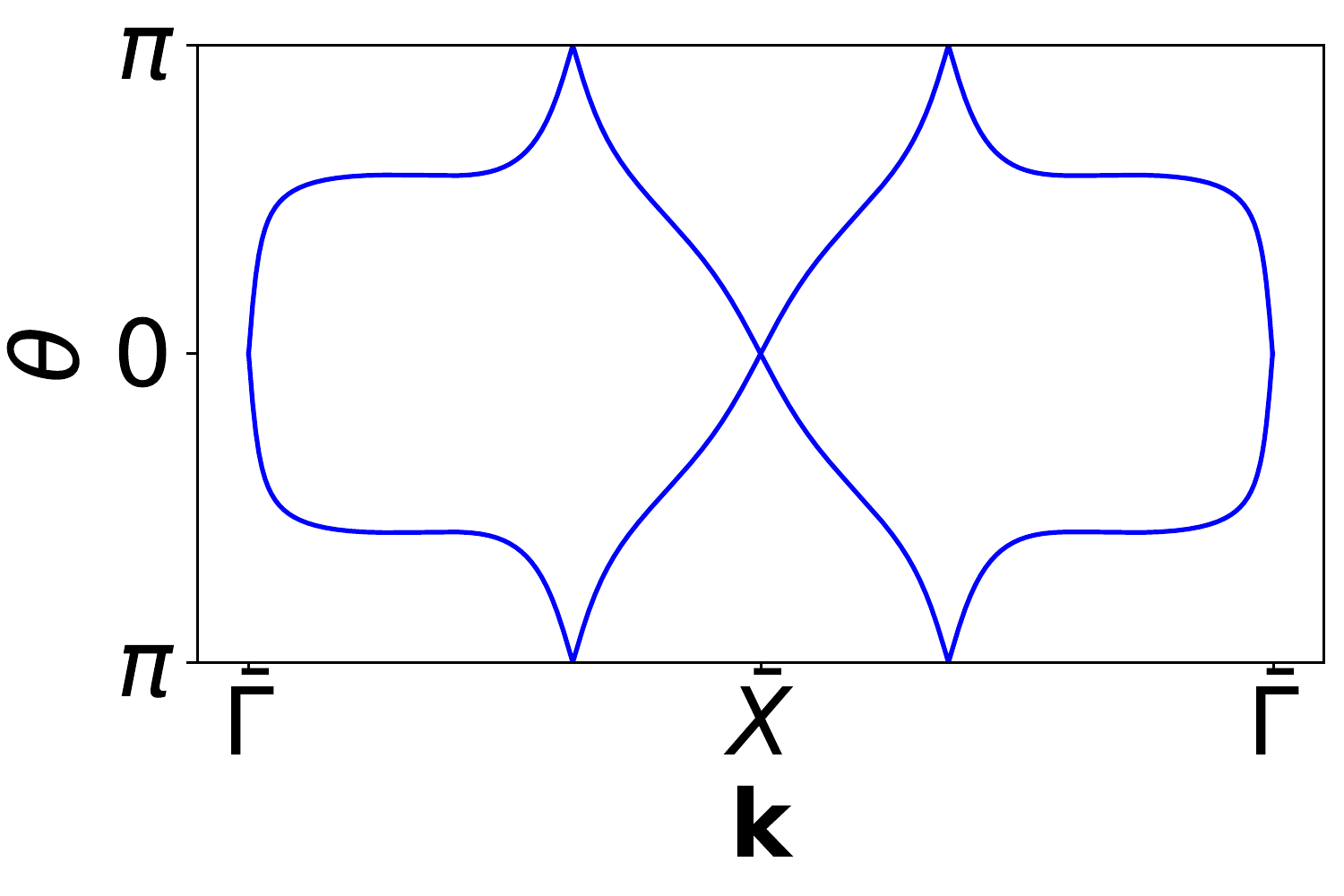}\label{fig:wilson}
}
\subfloat[]{
    \includegraphics[width=0.2\textwidth]{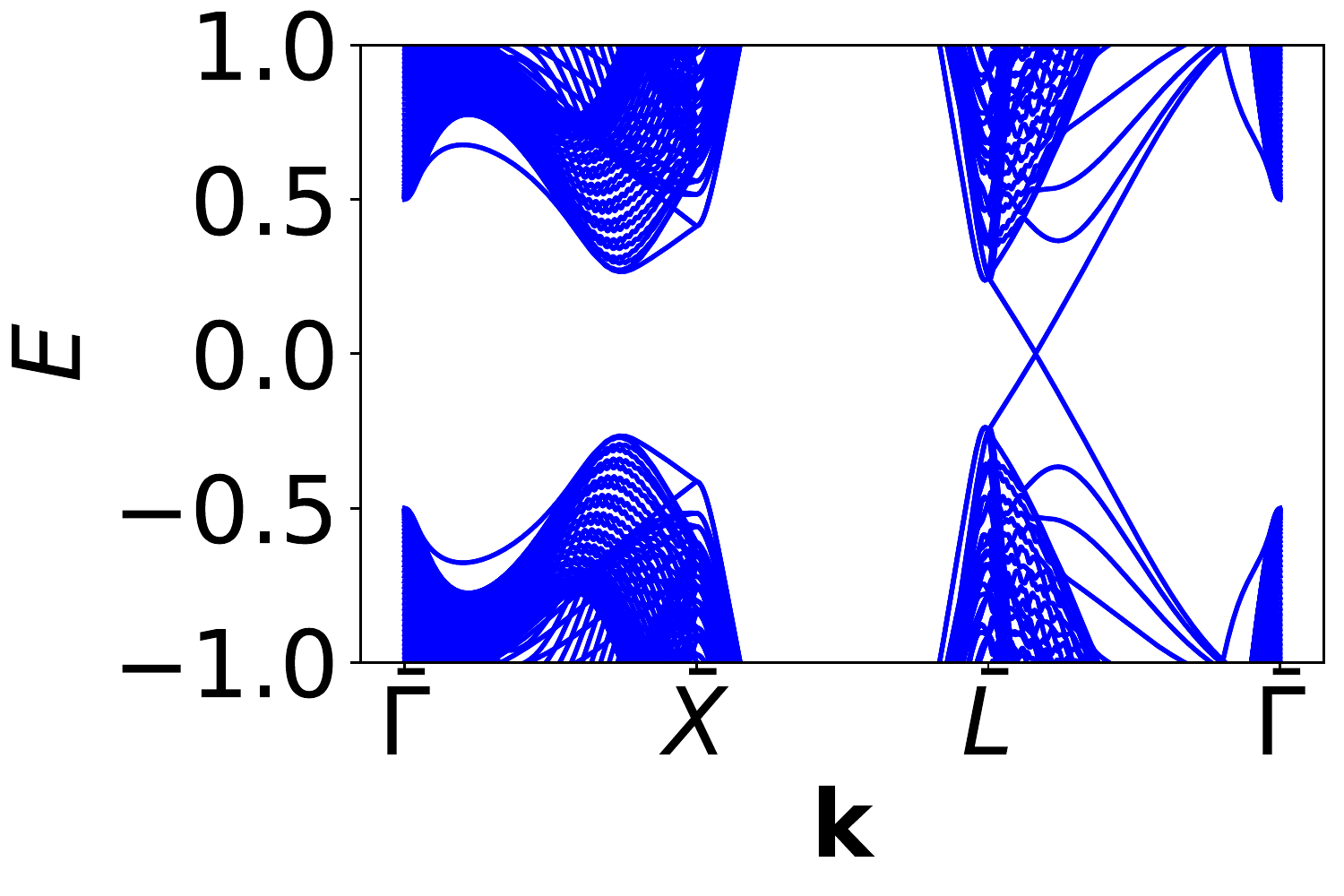}\label{fig:surface}
}
\caption{(a) and (b): Spectrum of the iFDB Hamiltonian given in Eq.~(\ref{eq:Hifdb}), with parameter values $T=\delta_2=0.5$, $\delta_1=0.1$. In (a) we take $m=0.3$, while in (b) we take $m=-0.3$. (c):  Wilson loop in the $k_2$ direction evaluated in the space of the lowest two bands of the model Eq.~(\ref{eq:Hifdb}) with $m<0$, as a function of $k_3$, with $k_1=0$. We see that the Wilson loop phases $\Theta$ wind twice around the circle $(-\pi,\pi]$. This implies that the model is in a topological phase with mirror Chern number $\nu_{m_{1\bar{1}0}}=2$. (d): Spectrum for the topological phase of the iFDB model for a $\mathbf{\hat{z}}$-normal slab. Note the mirror-symmetry protected Dirac fermion on the $\bar{\Gamma}-\bar{L}$ line.}
\end{figure}

To repair the symmetries, we seek hopping terms which vanish at the $L$ point, are nonvanishing everywhere else, and transform in the representation given by Eqs.~(\ref{eq:delta1}--\ref{eq:delta2}) along with
\begin{equation}
\Delta(C_{4x})=\exp(-i\frac{\pi}{4}\sigma_x).\label{eq:delta3}
\end{equation}
In this way, we will replicate the band inversion at $L$ in our symmetric model. Let us first fix the spin-orbit coupling term. Noting that the matrices $\tau_{x,y}\sigma_\mu$ transform in a pseudovector representation under roto-inversions, we need them to be multiplied by functions of $\mathbf{k}$ which are also pseudovectors. Combining this observation with time-reversal invariance and the boundary conditions on the Bloch Hamiltonian, we find that the simplest choice of SOC term which vanishes at $L$ is
\begin{equation}
H_{\mathrm{SOC}}=t\tau_y\sum_{\mu,\nu,\lambda=x,y,z}\left(\epsilon^{\mu\nu\lambda}\sigma_\mu\sin\frac{k_\nu}{2}\sin k_\lambda\right).
\end{equation}
For now, let us overlook the long (5-th nearest neighbor) range of this coupling in light of its mathematical simplicity. We may be tempted to take $m\tau_z+H_{\mathrm{SOC}}$ as our improved Hamiltonian, however $H_{\mathrm{SOC}}$ vanishes along the whole $\Gamma-L$ line, rather than just at the $L$ point. To remedy this, we can add two additional spin-independent hopping terms
\begin{equation}
H_{\mathrm{hop}}=\sum_{\mu=x,y,z}\left[\delta_1\tau_z\left(1+\cos k_\mu\right)+\delta_2\tau_x\cos\frac{k_\mu}{2}\right].
\end{equation}
We take for our full improved Hamiltonian
\begin{equation}
H_\mathrm{iFDB}=m\tau_z+H_{\mathrm{SOC}}+H_\mathrm{hop}\label{eq:Hifdb}.
\end{equation}
when $m=0$ with $t$ and $\delta_2$ nonzero, this model is gapless only at the $L$ point. for $m\neq 0$,  a nonzero $\delta_1$ ensures that spectrum of this Hamiltonian is gapped. We see that there is thus an insulator-to-insulator transition driven by an inversion of bands at the $L$ point, just as in the original $FDB$ model. We show the bulk spectrum for positive and negative values of $m$ in Fig.~\ref{fig:modelbulkspec1} and \ref{fig:modelbulkspec2}. 

\begin{table}[t]
\begin{tabular}{c|cccc}
EBR & $\Gamma$ & $X$ & $L$ & $W$ \\
\hline
$(\bar{E}_{1g})_{4a}\uparrow G$ & $\bar{\Gamma}_6$ & $\bar{X}_6$ & $\bar{L}_9$ & $\bar{W}_6$ \\
$(\bar{E}_{1g})_{4b}\uparrow G$ & $\bar{\Gamma}_6$ & $\bar{X}_6$ & $\bar{L}_8$ & $\bar{W}_7$
\end{tabular}
\caption{Little group irreps subduced by each of the elementary band representations in our model. The first column gives the name of the elementary band representation (EBR). The subsequent columns give the little group irreps at each of the high symmetry points.}\label{table:irreps}
\end{table}

\paragraph{Topological properties} Let us now examine the topological nature of this band-inversion transition. We will start by analyzing the band representations in the model following Refs.~\cite{NaturePaper,EBRTheoryPaper,bigmaterials}. Following the notation of the Bilbao Crystallographic Server\cite{Bilbao1,Bilbao2,Bilbao3,GroupTheoryPaper}, our model is induced from orbitals transforming under the $\bar{E}_{1g}$ representation of the point group $O_h$ on the $4a$ site, and a second set of orbitals transforming in the same representation on the $4b$ site. The four bands in our model thus transform under the composite band representation $(\bar{E}_{1g})_{4a}\uparrow G \oplus (\bar{E}_{1g})_{4b}\uparrow G$. These two elementary band representations subduce representations of the little group $G_\mathbf{k}$ at each of the high symmetry momenta $\Gamma,X,W,$ and $L$ in the Brillouin zone; we summarize these representations in Table~\ref{table:irreps} below. In the atomic limit of $H_{iFDB}$, we have $m\gg t,\delta_1,\delta_2$, and so the valence bands of our model transform in the $(\bar{E}_{4b}\uparrow G)$ elementary band representation, with occupied little group representations $\bar{\Gamma}_6,\bar{X}_6,\bar{W}_6,\bar{L}_9$; this can be checked explicitly using the representation $\Delta$ defined in Eqs.~(\ref{eq:delta1}--\ref{eq:delta2}) and (\ref{eq:delta3}) and taking into account the boundary conditions 
$|u_{n\mathbf{k}+n_i\mathbf{g}_i}\rangle=(\tau_z)^{\sum_i n_i}|u_n\mathbf{k}\rangle$,
where $\{\mathbf{g}_i\}$ is a basis for the reciprocal lattice, and $\{n_i\}$ are integers; this expresses the ``tight-binding gauge'' boundary conditions\cite{vanderbilt2018berry}. Upon inverting bands by taking $t\gtrsim \delta_2\gg \delta_1 > 0 > m, |m|\ll t$, we see that the occupied band irreps at $\Gamma, X$ and $W$ have not changed. At $L$ however, the wavefunctions now transform under the $\bar{L}_8$ irrep. Examining the full table of elementary band representations for $Fm\bar{3}m1'$\cite{progbandrep}, we find that the collection $(\bar{\Gamma}_6,\bar{X}_6,\bar{W}_6,\bar{L}_8)$ of occupied little group representations cannot be subduced by an integer sum \emph{or difference} of elementary band representations; we thus conclude that this phase of our model is a symmetry-indicated, stable topological crystalline insulator. Going further, we can attempt to express the irrep multiplicities in this model as a \emph{rational} sum of those in EBRs. Reading off the denominator of the rational coefficients, we find that symmetry-indicated TCIs in this space group are classified by an index $\nu\in \mathbb{Z}_8$. The index $\nu$ can be expressed as \cite{Po2017,khalaf,song2017}
\begin{equation}
\nu=\kappa_1-2\kappa_4 \mod 8,\label{eq:nudef}
\end{equation}
where $4\kappa_1$ is the sum of occupied band inversion eigenvalues, and $2\sqrt{2}\kappa_4$ is the sum of occupied band $IC_{4z}$ eigenvalues. Note that both $\kappa_1$ and $\kappa_4$ are integers\cite{khalaf}. For our model, we have $\nu=4$ in the topological phase, and $\nu=0$ in the trivial phase. As shown in Refs.~\onlinecite{khalaf,song2017}, a minimal model for a TCI with $\nu=4$ arises through a ``stacking'' (in Hilbert space) of four $\mathbb{Z}_2$ topological insulators. We thus expect to find protected gapless states on symmetric surfaces of this model, as we will discuss further below.

Furthermore, note that while the original FDB model is not $C_{4z}$ symmetric, it \emph{is} inversion symmetric. Thus, the sum of occupied band inversion eigenvalues $\kappa^{\mathrm{FDB}}_1$ can be computed for that model, and is given by
\begin{equation}
\kappa^{\mathrm{FDB}}_1=\begin{cases}
0, & m>0 \\
4, & m<0
\end{cases}
\end{equation}
Next, note that $IC_{4z}$ is not in the little group of any of the $L$ points. Therefore, in any symmetric completion of the FDB model the occupied band $IC_{4z}$ eigenvalues do not change as a function of $m$, and so neither does $\kappa_4$. Thus, in any symmetric completion of the FDB model, the index $\nu$ must change by $4$ as the sign of $m$ changes. Assuming additionally that the $m>0$ phase is connected to the (unobstructed\cite{NaturePaper}) atomic limit, we deduce that the band inversion in the FDB model becomes, when cubic symmetry is enforced, the transition between phases with $\nu=0$ and $\nu=4$.

As discussed in detail in Refs.~\onlinecite{song2017,khalaf}, the value of $\nu$ does not uniquely determine the topological phase of a system in space group $Fm\bar{3}m1'$. In particular, with $\nu=4$, there are two possible types of topological phase: the first has a mirror Chern number $\nu_{m_{z}}=4\mod 8$ associated with the $m_z$ mirror symmetry, while the second has both a mirror Chern number $\nu_{m_{1\bar{1}0}}=2\mod 8$ associated with the diagonal $\hat{\mathbf{x}}-\hat{\mathbf{y}}$ mirror symmetry, as well as a non-vanishing higher-order topological index. It is this latter phase which describes our current model.

We can make these statements more precise by examining the low-energy $\mathbf{k}\cdot\mathbf{p}$ theory for the topological transition in both $H_{FDB}$ and $H_{iFDB}$. Starting with the original FDB model, we find by expanding Eq.~(\ref{eq:fdb}) that
\begin{equation}
H_{\mathrm{FDB}}(L+\mathbf{k})\approx m\tau_z+\sum_{\mu=x,y,z}\frac{t}{2}\tau_y\sigma_\mu k_\mu.\label{eq:fdblowenergy}
\end{equation}
On the other hand, performing the same expansion of $H_{\mathrm{iFDB}}$, setting $\delta_2=t$, and defining the rotated coordinates $(k_a,k_b,k_c)$
and spin matrices
$(s_a,s_b,s_c)$
\footnote{Explicitly, these rotations are given by $\left(\begin{array}{c}
k_1 \\
k_2 \\
k_3 \\
\end{array}
\right)=\left(\begin{array}{ccc}
-\frac{1}{2\sqrt{6}} & -\frac{1}{2\sqrt{2}} & \frac{1}{\sqrt{3}} \\
-\frac{1}{2\sqrt{6}} & \frac{1}{2\sqrt{2}} & \frac{1}{\sqrt{3}} \\
\frac{1}{\sqrt{6}} & 0 & \frac{1}{\sqrt{3}} \\
\end{array}
\right)\left(\begin{array}{c}
k_a\\
k_b \\
k_c \\
\end{array}
\right)$ and 
$\left(\begin{array}{c}
s_x \\
s_y \\
s_z \\
\end{array}
\right)=\left(\begin{array}{ccc}
\frac{1}{\sqrt{2}} & \frac{\sqrt{3}+\sqrt{6}}{6} & \frac{\sqrt{2}-1}{2\sqrt{3}} \\
-\frac{1}{\sqrt{2}} & \frac{\sqrt{3}+\sqrt{6}}{6} & \frac{\sqrt{2}-1}{2\sqrt{3}} \\
0 & \frac{1}{\sqrt{6}}-\frac{1}{\sqrt{3}} & \frac{1}{\sqrt{6}}+\frac{1}{\sqrt{3}} \\
\end{array}
\right)\left(\begin{array}{c}
s_a\\
s_b \\
s_c \\
\end{array}
\right)$
} 
yields to \emph{quadratic} order
\begin{align}
H_{\mathrm{iFDB}}(L+\mathbf{k})&\approx[m+2\delta_1(k_a^2+k_b^2+k_z^2)]\tau_z\nonumber \\
&+\sqrt{3}\delta_2k_c\tau_x+2T\sqrt{3}\left(k_a\tau_ys_a+k_b\tau_ys_b\right)\label{eq:ifdblow}
\end{align}
Up to a choice of basis for the Dirac matrices, this is the BHZ model Hamiltonian for a topological insulator\cite{bernevig2006quantum,bernevigbook} -- note that because of our choice of boundary conditions and our expansion about the $L$ point, inversion symmetry is represented by $\Delta_L(I)=\tau_z$ 
in the $\mathbf{k}\cdot\mathbf{p}$ expansion. Eq.~(\ref{eq:ifdblow}) is also equivalent to Eq.~(\ref{eq:fdblowenergy}) if we take $\delta_2=2T=t/2$. Since there are $4$ L points, we see that this is a TCI rather than a TI transition. Furthermore, note that the plane $k_x=k_y$ corresponds to the plane $k_b=0$, and is invariant under $m_{1\bar{1}0}$; this symmetry is represented at the $L$ point by the matrix $\Delta_L(m_{1\bar{1}0})=\exp(i\pi/(2\sqrt{2})(s_a))$. Restricting $H_{\mathrm{iFDB}}$ to the mirror plane, we find that the Hamiltonian is block diagonal in the basis of $m_{1\bar{1}0}$ eigenstates, and describes a Chern insulator transition in each mirror subspace. Since there are two $L$ points in this mirror plane, we thus deduce, following Ref.~\onlinecite{Hsieh2012}, that this model corresponds to a mirror Chern insulator with $\nu_{m_{1\bar{1}0}}=2$. To verify this, we extract the mirror Chern number from the flow of hybrid Wannier charge centers\cite{taherinejad2014wannier,z2pack}, i.e. from the Wilson loop\cite{Yu11,Alexandradinata14}. We show in Fig.~\ref{fig:wilson} the $k_2$ directed Wilson loop for the occupied bands as a function of $k_3$. We see that in the $k_1=0$ plane the Wilson loop phases exhibit a nontrivial winding with winding number $2$; since in the FCC Brillouin zone this is the only mirror invariant plane\cite{Hsieh2012}, this signifies the mirror Chern number $\nu_{m_{1\bar{1}0}}=2$\cite{ArisCohomology}.

 As pointed out in Refs.~\onlinecite{song2017,hotis,khalaf}, models in space group $Fm\bar{3}m1'$ with $\nu=4$ and $\nu_{m_{1\bar{1}0}}=2$ also have a nontrivial higher order index $\delta_{S_4}=1$ protected by fourfold rotoinversion symmetry. This phase has four Dirac cones on a $\mathbf{\hat{z}}$-normal surface, thus explaining the four Dirac fermions on an antiphase boundary found in Refs.\cite{fradkin1,fradkin2,fradkin3}: At the level of the tight-binding model, an antiphase domain wall is simply a boundary between the trivial ($m>0$) and topological ($m<0$) phase of the model. These domain wall fermions are indeed of topological origin, and are symmetry protected in any symmetric extension of the FDB model. We show the spectrum of the topological phase of Eq.~(\ref{eq:Hifdb}) on a $\mathbf{\hat{z}}$-normal slab in Fig.~\ref{fig:surface}

Finally, note that the original low-energy FDB Hamiltonian Eq.~(\ref{eq:fdblowenergy}) at linear order has an accidental mirror symmetry
$\tilde{\Delta}_L(m_{1\bar{1}0})=\tau_z\exp(i\pi/(2\sqrt{2})(s_x-s_y))$,
which leaves the plane $k_x=k_y$ invariant. While this symmetry is inconsistent with the inversion symmetry of the full model (and so will be broken by crystal-symmetry preserving perturbations, including higher-order terms in the $\mathbf{k}\cdot\mathbf{p}$ expansion), it explains why the authors of Ref.~\onlinecite{fradkin1,fradkin2,fradkin3} were able to find domain wall fermions in their model.

\begin{figure}[t]
\includegraphics[width=0.45\textwidth]{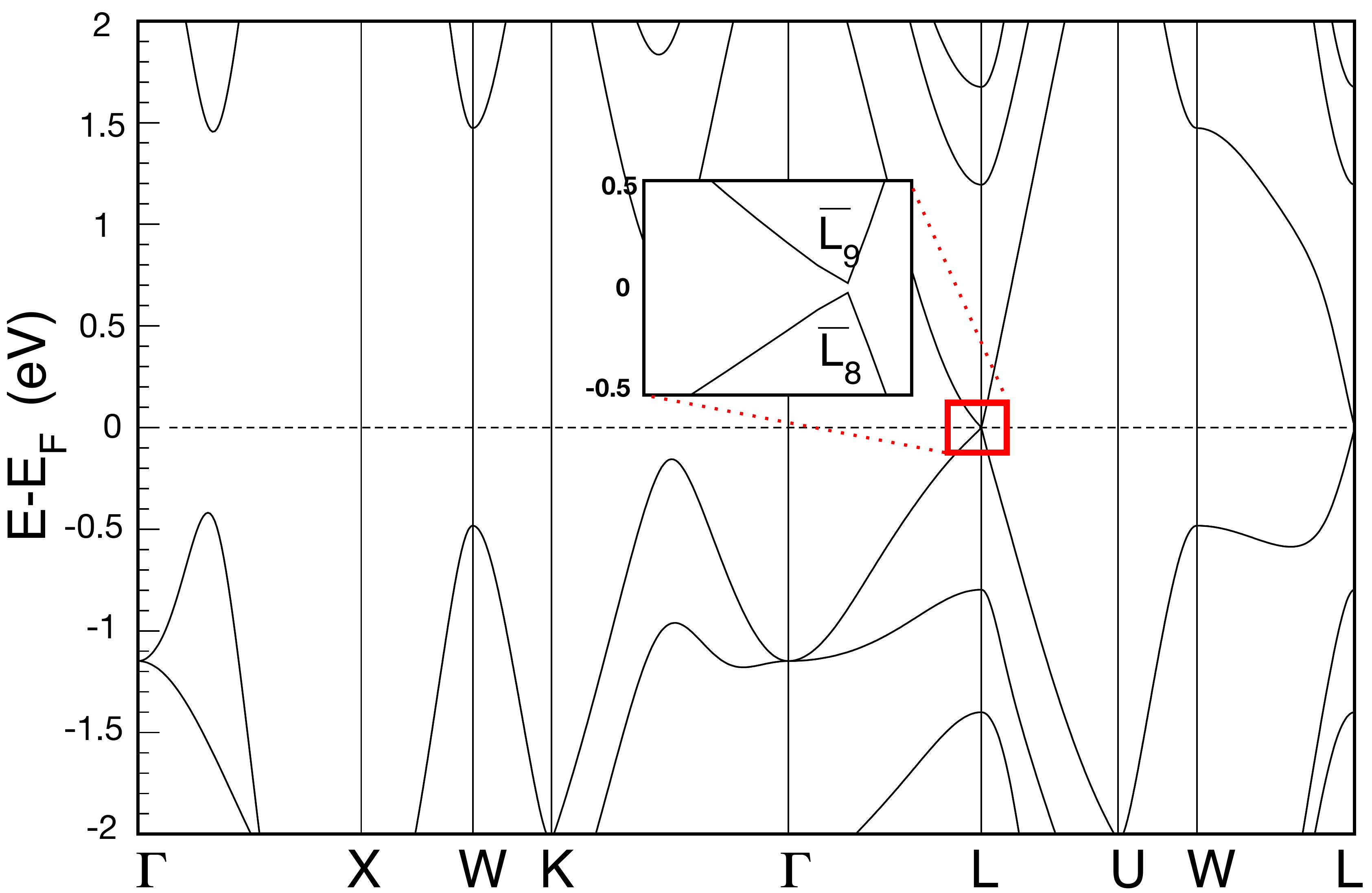}
\caption{Bulk band structure of PbTe, calculated using the structure reported in Ref.~\onlinecite{pbtestruct}. The inset shows the small gap and band inversion at the $L$ point.}\label{fig:pbtebulkbands}
\end{figure}

\begin{figure}[t]
\includegraphics[width=0.4\textwidth]{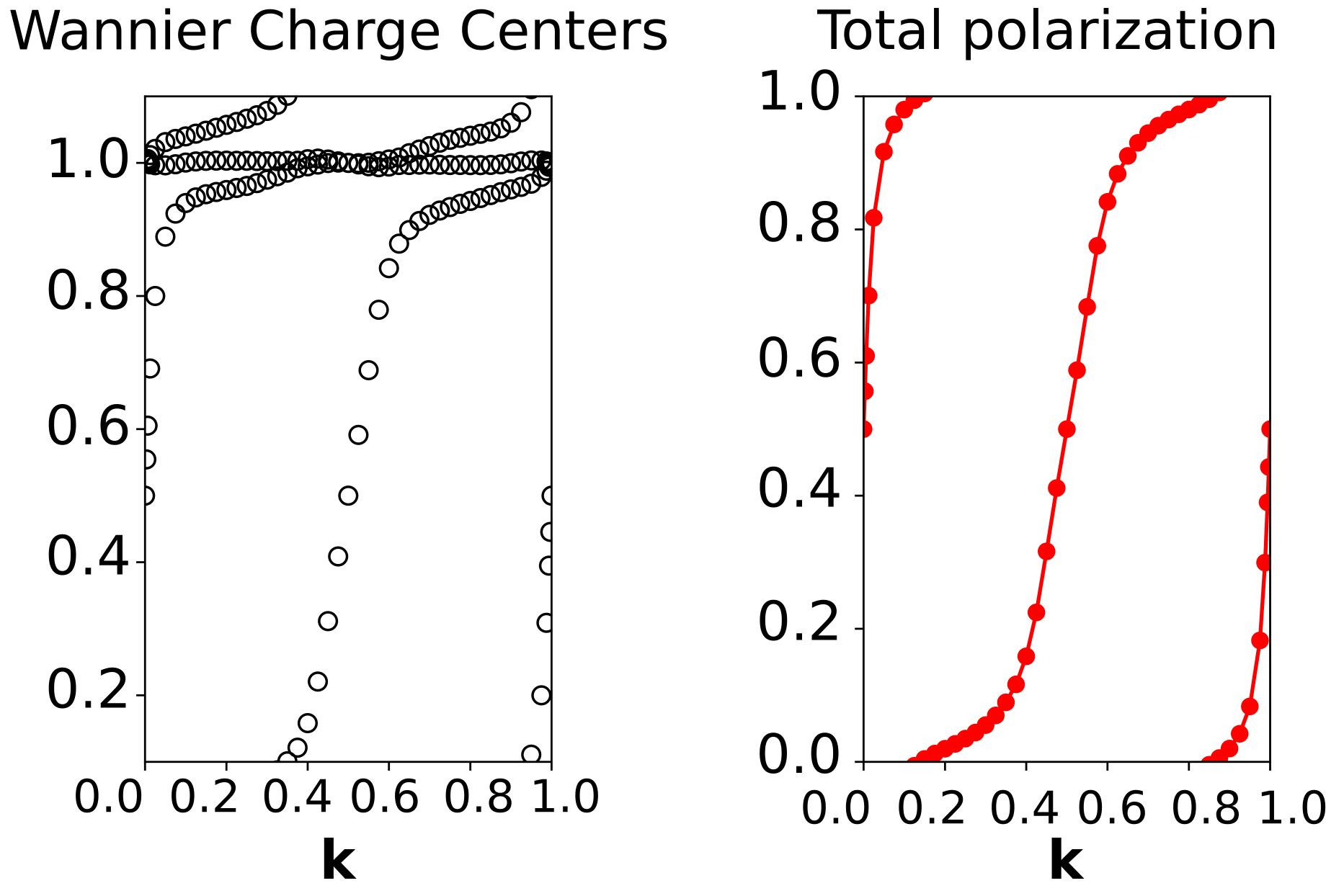}
\caption{Hybrid Wannier charge centers for the $+i$ mirror subspace of the occupied bands of PbTe in the $m_{110}$-invariant plane. The left shows the raw Wannier centers (Wilson loop eigenvalues), while the right shows their sum. Since the sum ``winds'' twice across the unit cell, we deduce that the mirror chern number $\nu_{m_{1\bar{1}0}}=2$.}\label{fig:pbteloops}
\end{figure}

\paragraph{Ab-initio results} In the previous sections, we have seen how symmetrized completions of the FDB model of PbTe yield higher order topological and mirror Chern insulators. Under certain conditions, the band structure for realistic PbTe as computed with ab-initio methods realizes this same $\nu=4$ symmetry-indicated phase. This can be seen by analyzing the material catalogues of Refs.~\onlinecite{bigmaterials,bigmaterials-china}, which both report the value of $\nu=4$ for PbTe. For confirmation, we have recomputed the ab-initio band structure of PbTe using {{Density Functional Theory (DFT)\cite{Hohenberg-PR64,Kohn-PR65} as implemented in the Vienna Ab initio Simulation Package (VASP)\cite{Kresse199615,PhysRevB.48.13115}. We use the structural parameters as reported in Ref.~\onlinecite{pbtestruct}. The interaction between ion cores and valence electrons was treated by the projector augmented-wave method\cite{vaspPaw}, the generalized gradient approximation (GGA) for the exchange-correlation potential with the Perdew-Burke-Ernkzerhof for solids parameterization~\cite{PBE} and spin-orbit coupling was taken into account by the second variation method\cite{PhysRevB.62.11556}. A Monkhorst-Pack centered at $\Gamma$
k-point grid of (11$\times$11$\times$11) for reciprocal space integration and 500 eV energy cutoff of the plane-wave expansion have been used}}. We show the band structure in Fig.~\ref{fig:pbtebulkbands}, with an inset highlighting the rather small gap at $L$. Employing the VASPtoTrace tool\cite{progvasptotrace,bigmaterials}, we compute the little group irreps of the occupied bands at the high symmetry points, shown in Table~\ref{table:abinit}; we give the irreps of SnTe as well for comparison. By using Eq.~(\ref{eq:nudef}), we see that $\nu=4$ for both SnTe and PbTe. Note, in fact, that the irrep labels for SnTe and PbTe differ only in a shift of the origin of the system by $(\half\half\half)$
Furthermore, the topological transition to $\nu=4$ in the realistic material is driven by a band inversion of the irreps at $L$ relative to $W$, just as in the FDB model. To fully determine the topological phase, we evaluate the mirror Chern number of the occupied bands using Z2Pack\cite{z2pack}. We find a mirror Chern number $\nu_{m_{110}}=\nu_{m_{1\bar{1}0}}=2$, just as in the iFDB model\footnote{These two mirror chern numbers are equal due to the cubic symmetry of the system.} (See Fig.~\ref{fig:pbteloops}).

However, it is well-accepted that the mirror Chern number $\nu_{m_{1\bar{1}0}}$ in PbTe is zero under ambient experimental conditions, while it is $2$ for SnTe\cite{Hsieh2012}. To reconcile this with the nontrivial $\nu=4$ topological index, we note that in addition to the structure used for the ab initio calculations here and in Refs.~\onlinecite{bigmaterials,bigmaterials-china}, PbTe has 41 other entries in the ICSD\cite{ICSD} in the space group $Fm\bar{3}m1'$\cite{bigmaterials}. A DFT analysis of other structures (for instance, the structure reported in Ref.~\onlinecite{STERNBERG1982364}) yields a trivial index $\nu=0$ due to a band de-inversion at $L$, in agreement with the experimental findings. This highlights the fact that for small band gap insulators, one must be cautious in extracting the band topology from ab initio calculations; for PbTe in particular, the failure of semilocal DFT to correctly produce the (correct sign of the) experimental band gap in certain cases has been noted previously\cite{Kressepbte}. 

To investigate this systematically, we have computed the band structures and topological index $\nu$ for all $42$ entries of PbTe in the ICSD, using the PBE functional. The input parameters for these compounds differ only in the reported lattice constant $a_0$, which range between $6.157$\AA~and $6.543$\AA. For the six reported structures with $a_0\leq 6.44$\AA, PBE predicts $\nu=4$; for the remaining with larger lattice constants we find $\nu=0$. In the Supplementary Material we give a table summarizing our DFT calculations\footnote{See Supplemental Material at \url{https://journals.aps.org/prmaterials/supplemental/10.1103/PhysRevMaterials.3.041202/suppmat.pdf} for the topological index for all reported entries of PbTe.}. Taken optimistically, This shows that PbTe is very close to a topological phase transition, which may be tunable as a function of external parameters such as hydrostatic pressure.

\begin{table}[t]
\begin{tabular}{c|c|c}
$\mathbf{k}$ & SnTe & PbTe \\
\hline
$\Gamma$ &$\bar{\Gamma}_6,\bar{\Gamma}_6,\bar{\Gamma}_8,\bar{\Gamma}_{11}$ & $\bar{\Gamma}_6,\bar{\Gamma}_6,\bar{\Gamma}_8,\bar{\Gamma}_{11}$ \\
$X$ & $\bar{X}_6,\bar{X}_6,\bar{X}_8,\bar{X}_8,\bar{X}_9$ & $\bar{X}_6,\bar{X}_6,\bar{X}_8,\bar{X}_8,\bar{X}_9$ \\
$L$ & $\bar{L}_9,\bar{L}_8,\bar{L}_8,\bar{L}_4\bar{L}_5,\bar{L}_9$ & $\bar{L}_8,\bar{L}_9,\bar{L}_9,\bar{L}_6\bar{L}_7,\bar{L}_8$ \\
$W$ & $\bar{W}_6,\bar{W}_7,\bar{W}_7,\bar{W}_6,\bar{W}_7$ & $\bar{W}_7,\bar{W}_6,\bar{W}_7,\bar{W}_6,\bar{W}_6$ \\
\end{tabular}
\caption{Occupied band irreps for SnTe and PbTe at the high symmetry points. Irreps are listed in order of increasing energy, i.e. those states closest to the fermi level appear at the end of the list. Note that the irreps at $\Gamma$ and $X$ are identical for the two materials.}\label{table:abinit}
\end{table}

\paragraph{Conclusion} We have revisited the effective model of PbTe as presented in Refs.~\onlinecite{fradkin1,fradkin2,fradkin3}. We have shown that the domain wall fermions in the FDB model, long derided as nontopological, are signatures of the topological surface states present in any symmetric completion of the model, protected by mirror and fourfold rotoinversion symmetries. Furthermore, we show that ab initio calculations reveal that some of the reported structures of realistic PbTe are in this same symmetry-indicated class of materials, at least within the GGA. This shows that PbTe is an ideal platform for exploring structurally-tunable topological behavior. Finally, while within the context of our effective model there is no difference between an antiphase domain wall and a domain wall with the vacuum, this is not necessarily true in a more realistic system. Given the recent focus on defect response of higher-order topological insulators\cite{slager2015,queiroz2018partial,hingeprep,benalcazar2018quantization}, it would be interesting to examine this more carefully for both SnTe and PbTe structural variations in future work.\cite{frankprep,inigo}
\begin{acknowledgments}
\paragraph{Acknowledgements:} BB and MGV thank E. Fradkin for helpful discussions, and for pointing out Refs.~\cite{fradkin2,fradkin3}. BB additionally acknowledges discussions with B.~J. Wieder. This work was initiated at the Aspen Center for Physics, which is supported by National Science Foundation grant PHY-1607611. MGV acknowledges the IS2016-75862-P national project of the Spanish MINECO. 
\end{acknowledgments}

\bibliography{refs}
\end{document}